\documentclass[aps,prl,showpacs,superscriptaddress]{revtex4}
\usepackage{amsmath}
\usepackage{amssymb}
\usepackage{epsfig}
\usepackage{color}
\usepackage{graphicx,amsmath}

\begin{document}
\newcommand{\beq}{\begin{equation}}
\newcommand{\eeq}{\end{equation}}

\title{Theory of fermion condensation as  an analog of the liquid-drop theory of atomic nuclei}
\author{V.~A. Khodel}
\affiliation{NRC Kurchatov Institute, Moscow, 123182, Russia}
\affiliation{McDonnell Center for the Space Sciences \& Department
of Physics, Washington University, St.~Louis, MO 63130, USA}
\begin{abstract}
Employing the duality between  the momentum distribution $n({\bf
p})$ and  density distribution $\rho({\bf r})$, problems of theory
of  systems with flat bands, pinned   to the Fermi surface, are
discussed. We propose that the Lifshitz topological   phase
transition associated with  the formation of  additional pockets of
the Fermi surface   is the precursor of fermion condensation.
\end{abstract}

\maketitle

A theory of strongly correlated Fermi systems, called   theory of
fermion condensation, was created more than 20 years ago
\cite{ks,vol,noz},  (for recent articles, see e.g.
\cite{vollhardt,prb2008,lee,shagrep,vol2011,mig100,an2012,kats}). A
key feature of the  phenomenon of fermion condensation is related
to {\it swelling  the Fermi surface}: in  momentum space, there
exists a domain $\Omega$, in which the $T=0$ single-particle
spectrum  $\epsilon({\bf p})$ becomes completely flat: \beq
\epsilon({\bf p})=\mu,  \quad {\bf p}\in \Omega  , \label{eq1} \eeq
where $\mu$ stands for the chemical potential of the system. The
totality of  these  states is called the fermion condensate (FC),
because the $T=0$ density of states of Fermi   systems with a FC
diverges in the same way $\propto \delta (\varepsilon-\mu)$
\cite{prb2008} as the density of states of systems with a Bose
condensate.

On the other hand, the spectrum $\epsilon( {\bf p})$ is known to be
a variational derivative of the ground-state energy functional
$E(n)$ with respect to  the momentum distribution $n( {\bf p})$. As
a result, in three-dimensional homogeneous Fermi liquid (FL),  the
basic Eq. (\ref{eq1}) of theory of fermion condensation  can be
rewritten as a variational condition \cite{ks} \beq {\delta
E(n)\over \delta n( p)}=\mu , \quad  p_{min}<p<p_{max}  ,
\label{var} \eeq determining the FC momentum distribution $n_*( p)$
that  changes continuously between   0 and 1 inside a momentum
interval $p_{min}<p<p_{max}$. The FC boundaries $p_{min}$ and
$p_{max}$ depend on normalization condition. In   the
strong-coupling limit where the Pauli restriction $n(p)\leq 1$ is
automatically  met, it reduces to \beq \rho=\int\limits_0^{p_{max}}
n_*(p)    {p^2_1dp_1\over \pi^2}. \label{norfc} \eeq Another key
feature of fermion condensation is that its  onset is associated
with  a rearrangement  of the topological structure of the ground
state. In contrast to conventional FLs, which possess the integer
topological charge, beyond the FC transition point its value
becomes half-integer, and this topological feature is robust to
variations of input parameters \cite{vol}.

The solutions $n_*( p)$ of Eq.(\ref{var}) are  found  solely in
analytically solvable models  where the effective  energy
functional $E(n)$ has a Hartree-like form \beq E(n)=\sum
\epsilon^0_p n({\bf p})+{1\over 2}\sum\sum U({\bf p}-{\bf p}_1)
n({\bf p}) n({\bf p}_1) , \label{hef} \eeq with the interaction
$U(q)$, possessing a singularity at $q=|{\bf p}-{\bf p}_1|\to 0$,
(see, e.g. \cite{ks,noz,physrep}). Furthermore,    flat bands
pinned to the Fermi surface never emerge in iterative  numerical
calculations  of Landau equation \cite{lan} \beq {\partial
\epsilon( {\bf p})\over \partial {\bf p}}={{\bf p}\over M}+\int
f({\bf p},{\bf p}_1) {\partial n({\bf p}_1)\over \partial {\bf
p}_1}d \upsilon_1 , \label{sp} \eeq (with $d\upsilon=2d^3p/
(2\pi)^3$),  except for the models where the interaction functions
$f$ are singular, as mentioned above (see e.g. \cite{zb,prb2008}).

Since numerical calculations of Eq.(\ref{sp}) are practically the
sole source  of information on the momentum distribution $n(p)$
beyond the point of a topological rearrangement of the Landau
state, concerns were repeatedly voiced that in Fermi systems with
interaction functions regular in momentum space, nontrivial
solutions   of Eq.(\ref{var}) are  nonexistent, allegedly due to
violation of analytic properties, inherent in relevant equations of
FL theory.   Therefore beyond the critical point, breeding
additional Lifshitz pockets of the Fermi surface with FL occupation
numbers $0,1$ \cite{lifshitz,zb} is the single possible
alternative.

However, as we will see, this is not the case. In strongly
correlated Fermi systems,  phases  with the FC do exist
irrespective of analytic properties  of the input. The  analogous
situation  takes place  in the well-studied liquid-drop   theory
(LDT) of atomic nuclei.  Although this theory is formulated in
coordinate space, rather than in  momentum space, both the spaces
are equivalent from  the mathematical standpoint, and the momentum
distribution $n({\bf p})$ is dual to the density $\rho({\bf r})$.
In condensed matter theory,  this circumstance  is employed already
for more than 20 years, (see e.g. \cite{ks,lee,kane}), and
therefore it is instructive to exploit the duality  to clarify the
disputed topic.

LDT states are known to emerge as nontrivial solutions of
variational equation \beq {\delta E(\rho)\over \delta \rho({\bf
r})}=\mu , \label{varr} \eeq supplemented with    normalization
condition
  \beq
  N=\int \rho({\bf r})d{\bf r}  ,
  \label{norm}
  \eeq
where  $N$ stands for a {\it finite} particle number. Obviously,
these two equations are equivalent to Eqs. (\ref{var}) and
(\ref{norfc}).

To be more specific, the  LDT  energy functional $E(\rho)$ consists
of the kinetic energy $\tau(\rho)$, proportional to
$\rho^{5/3}({\bf r})$, and an interaction term,  often written in
the Skirme-like form \beq
  E_{int}(\rho)={1\over 2}\int\int V({\bf r}_1,{\bf r}_2) \rho({\bf r}_1)\rho({\bf r}_2) d{\bf r}_1d {\bf r}_2 +
{1\over 3}\int\int\int W({\bf r}_1,{\bf r}_2,{\bf r}_3) \rho({\bf
r}_1)\rho({\bf r}_2) \rho({\bf r}_3) d{\bf r}_1d {\bf r}_2d{\bf
r}_3, \label{fenld} \eeq with interaction potentials $V$ and $W$,
vanishing at large distances between interacting particles. For
completeness, this expression should have being supplemented with
an external field term, say, a gravitational one, however, in the
case under consideration, it is of little  interest. In  spherical
systems Eq.(\ref{varr}) reads \beq \mu={p^2_F(\rho(r))\over
2M}+\int V({\bf r},{\bf r}_1) \rho( r_1) d{\bf r}_1 +\int\int
W({\bf r},{\bf r}_1,{\bf r}_2)
 \rho(r_1)\rho(r_2) d{\bf r}_1d {\bf r}_2,
\label{vare} \eeq with $p_F(\rho)=(3\pi^2\rho)^{1/3}$.

Since $N$ is finite,  the true density $\rho( r)$ must  rapidly
fall at $r\to\infty$. This fact is of paramount importance  for the
analytic structure of solutions of Eq.(\ref{vare}), depending on
whether $\mu$ equals to 0 or not.  In the case $\mu=0$,
Eq.(\ref{vare}) is met at any distance, implying that the solution
$\rho(r)$  remains finite,  analogously to the case of the
Rutherford electron atom in the Thomas-Fermi (TF) model \cite{LL}.
Contrariwise, at $\mu<0$ Eq.(\ref{vare}) cannot be satisfied at
$r\to\infty$, because the l.h.s. of this equation {\it remains
finite}, while the  r.h.s.  {\it comes to  nought}. This compels us
to conclude that Eq.(\ref{vare}) holds only at   $r<R$,  beyond
which $\rho( r)$ vanishes {\it identically}.  The situation is
reminiscent of the situation that takes place in electrically
charged atoms, which  have a {\it finite TF radius} R \cite{LL}.
Noteworthy, the kinetic term $p^2_F(\rho)/2M$  possesses the same
analytic properties as the density $\rho$ itself, and therefore its
presence on the r.h.s. of Eq.(\ref{vare}) or absence in the limited
case $M\to \infty$ has no impact on this conclusion.

Explicitly one has
\begin{eqnarray}
 \mu&=& {p^2_F(\rho(r))\over 2M}+\int V({\bf r},{\bf r}_1) \rho( r_1) d{\bf r}_1 +\int\int
W({\bf r},{\bf r}_1,{\bf r}_2) \rho(r_1)\rho( r_2) d{\bf r}_1d {\bf r}_2 , \quad   r<R, \nonumber\\
 \rho( r)&\equiv &0, \quad r>R  .
 \label{set1}
 \end{eqnarray}

To be certain that a nontrivial solution of the variational
condition (\ref{varr}) exists, it is sufficient to verify that the
above energy functional   attains minimum on a simple  class of
variational functions. If so,  nontrivial solutions of
Eq.(\ref{set1}) do exist, independent of  the presence or absence
of singularities in  the interaction potentials $V$ and $W$.

The  derivative $\delta E(\rho)/\delta \rho( r)$ is  compared with
the single-particle energy $\epsilon( r)$, while the derivative
$\partial \epsilon( r)/\partial r$ determines the force, acting on
a probe particle. Definitely, this force {\it identically vanishes}
in the nuclear interior, while outside the nucleus it reduces to
the Coulomb  term, produced by the proton component. Bearing in
mind the duality,  it is instructive to rewrite the spectrum
$\epsilon(r)$ itself in the form
\begin{eqnarray}
  \epsilon( r)&\equiv&\mu ,  \quad r<R ,\nonumber\\
   \epsilon( r)&=&\int V({\bf r},{\bf r}_1) \rho({\bf r}_1) d{\bf r}_1 +\int\int W({\bf r},{\bf r}_1,{\bf r}_2)
 \rho({\bf r}_1)\rho({\bf r}_2) d{\bf r}_1d {\bf r}_2 , \quad r>R.
 \label{epsr}
 \end{eqnarray}
Thereby the function $\partial \epsilon( r)/\partial r$ is {\it not
analytic}.

Now let us turn to the problem of fermion condensation. Remembering
that the quantity $n(p)$ is dual to the density $\rho(r)$, the
energy functional of theory of  fermion condensation  can be
written in the form, analogous  to that  in the LDT,
 \beq
  E(n)=\int  {p^2\over 2M} n(p)d\upsilon+{1\over 2}\int\int V({\bf p}_1,{\bf p}_2) n( p_1)n(p_2) d\upsilon_1d\upsilon_2 +
{1\over 3}\int\int\int W({\bf p}_1,{\bf p}_2,{\bf p}_3) n(p_1)n( p_2) n( p_3) d\upsilon_1d\upsilon_2d\upsilon_3  ,
\label{fenfc}
  \eeq
with the single difference:  the first kinetic-energy term on the
r.h.s. of this equation plays the part of an external harmonic
field.    The corresponding variational equation (\ref{var}) has
the   explicit form \beq
  \mu={p^2\over 2M}+\int V({\bf p},{\bf p}_1) n_*(p_1)d\upsilon_1 +\int\int W({\bf p},{\bf p}_1,{\bf p}_2)n_*(p_1)n_*(p_2) d\upsilon_1d\upsilon_2 , \quad p_{min}<p<p_{max}  .
 \label{varmu}
 \eeq
supplemented with  normalization condition (\ref{norfc}). In the
strong-coupling limit one has $p_{min}=0$, and therefore   $n(p)=0$
at $p>p_{max}$, otherwise   $n(p)=1$ at $p<p_{min}$, due to the
Pauli restriction. Evidently, the quasiparticle momentum
distribution  $n(p)$ of the FC problem does not possess properties
inherent in analytic functions. Behavior of  the group velocity
$\partial \epsilon(p)/\partial p $, which  identically vanishes  in
the FC region, is analogous to that of  the derivative $\partial
\epsilon(r)/\partial r$, vanishing in the nuclear interior. This
implies that the group velocity $\partial \epsilon(p,n_*)/\partial
p $ is not analytic. Definitely, the same is valid  for the
single-particle spectrum $\epsilon(p,n_*)$ of the FC problem, given
by equation
    \begin{eqnarray}
 \epsilon(p,n_*)&\equiv &\mu , \quad p<p_{max}  , \nonumber\\
  \epsilon(p,n_*)&=&{p^2\over 2M}+\int V({\bf p},{\bf p}_1) n_*(p_1) d\upsilon_1  +\int\int W({\bf p},{\bf p}_1,{\bf p}_2)n_*(p_1)n_*(p_2) d\upsilon_1 d\upsilon_2  ,\quad p>p_{max} ,
  \label{enfc}
  \end{eqnarray}
analogous to Eq.(\ref{epsr}).  Within the context of the Landau
approach to theory of Fermi liquid, equation appropriate for
numerical calculations of the FC structure has the closed form \beq
0={p\over M} +{1\over 3}\int f_1(p,p_1){\partial n_*(p_1)\over
\partial p_1} {p^2_1dp_1\over \pi^2}  ,\quad p<p_{max},
\label{fcf1} \eeq with $f_1$ standing for the first Legendre
harmonic of the Landau interaction function $f({\bf p},{\bf p}_1)$,
which is a phenomenological input.

It should be emphasized that Eqs.(\ref{sp}) and (\ref{fcf1}) are
{\it different}. In the strong-coupling limit where the Pauli
restriction $n(p)<1$  is of minor importance, the domain of the
Lifshitz phase diagram,  occupied by  the phase with the FC,
coincides with  the domain where Eqs. (\ref{var}) and/or
(\ref{fcf1}) have nontrivial solution $n_*(p)$. In this limit, a
different solution $n_L$, that satisfies   Eq.(\ref{sp}) and
consists of a set of additional Lifshitz  pockets of the Fermi
surface, is of little interest, because   the FC distribution
$n_*(p)$ {\it meets variational condition} (\ref{var}), while its
Lifshitz  counterpart $n_L(p)$, not. With {\it decreasing} the
coupling constant, the role of the Pauli principle {\it grows}, so
that the Pauli restriction is  eventually violated at some critical
point where $n_*(p)$ turns out to be in excess of unity.  A
plausible scenario for this part of the Lifshitz phase diagram is
associated with  the formation of a set of Lifshitz pockets. As the
coupling constant  decreases,   the Lifshitz pockets presumably
coexist  with FC islands until the very last FC island disappears,
and   the true momentum distribution $n(p)$ then obeys
Eq.(\ref{sp}).

In conclusion,  beyond the critical point where the  Landau
quasiparticle momentum distribution $n_F(p)=\theta(p_F-p)$ loses
its stability, the structure of the true momentum distribution
$n(p)$ evaluated on the base of Eq.(\ref{sp}) is in agreement  with
a conventional scenario\cite{lifshitz,zb,shagp,schofield,tai1,tai2}
for a topological rearrangement of $n(p)$ that reduces  to  the
generation of a finite number of additional  pockets of the Fermi
surface.  As the coupling constant increases, Eq.(\ref{sp}) and the
conventional scenario hold until   the onset of fermion
condensation, associated with the emergence  of   nontrivial
solutions of Eq.(\ref{fcf1}) provided these solutions  do not
violate the Pauli principle. Thereby the Lifshitz topological
transition \cite{lifshitz} is the {\it precursor} of fermion
condensation, and in the Lifshitz phase diagram of correlated Fermi
systems, the  phase with additional  pockets of the Fermi surface
separates  the conventional Landau phase from the phase with the
fermion condensate.

I thank V. Shaginyan and M. Zverev for fruitful discussions. This
work is supported by RFBR Grants 11-02-00467 and 13-02-00085.

\end{document}